\documentclass{aa}  

\usepackage{graphicx}
\usepackage{txfonts}
\usepackage[]{hyperref}
\usepackage{longtable}
\usepackage{amsmath}	
\usepackage{amssymb}	
\usepackage{multicol}        
\usepackage{pdflscape}	

\usepackage[T1]{fontenc}
\usepackage{ae,aecompl}

\usepackage{newtxtext,newtxmath}
\begin{document} 

\title{The physical properties of T Pyx as measured by MUSE}
\subtitle{I. The geometrical distribution of the ejecta and the distance to the remnant}

\author{L. Izzo\inst{1,2},
L. Pasquini\inst{3},
E. Aydi\inst{4},
M. Della Valle\inst{1},
R. Gilmozzi\inst{3},
E. A. Harvey\inst{5},
P. Molaro\inst{6,7}\\
M. Otulakowska-Hypka\inst{8},
P. Selvelli\inst{6},
C. C. Th\"one\inst{9},
R. Williams\inst{10,11}\\
}

   \authorrunning{Izzo et al.}
\institute{INAF, Osservatorio Astronomico di Capodimonte, Salita Moiariello 16, I-80131 Napoli, Italy\\
    \email{luca.izzo@inaf.it}
    \and
    DARK, Niels Bohr Institute, University of Copenhagen, Jagtvej 128, 2200 Copenhagen, Denmark
    \and
    European Southern Observatory, Karl Schwarzschild-Str. 2, 85748 Garching, Germany 
    \and
    Center for Data Intensive and Time Domain Astronomy, Department of Physics and Astronomy, Michigan State University, East Lansing, MI 48824, USA
    \and
    UK Astronomy Technology Centre, Royal Observatory Edinburgh, EH9 3HJ, Edinburgh, UK
    \and
    INAF-Osservatorio Astronomico di Trieste, Via G.B. Tiepolo 11, I-34143 Trieste, Italy 
    \and
    Institute of Fundamental Physics of the Universe, Via Beirut 2, Miramare, I-34151 Trieste, Italy
    \and 
    Astronomical Observatory Institute, Faculty of Physics, Adam Mickiewicz University, Słoneczna 36, 60-286 Poznań, Poland
    \and 
    Astronomical Institute, Czech Academy of Sciences, Fri\v cova 298, Ond\v rejov, Czech Republic  
    \and
    Department of Astronomy and Astrophysics, University of California, Santa Cruz, 1156 High Street, Santa Cruz, CA 95064, USA
    \and
    Space Telescope Science Institute, 3700 San Martin Drive, Baltimore, MD 21218, USA
}

\date{}

\abstract
{T Pyx is one of the most enigmatic recurrent novae, and it has been proposed as a potential Galactic type-Ia supernova progenitor.}
{Using spatially-resolved data obtained with MUSE, we characterized the geometrical distribution of the material expelled in previous outbursts surrounding the white dwarf progenitor.}
{We used a 3D model for the ejecta to determine the geometric distribution of the extended remnant. We have also calculated the nebular parallax distance ($d = 3.55 \pm 0.77$ kpc) based on the measured velocity and spatial shift of the 2011 bipolar ejecta. These findings confirm previous results, including data from the GAIA mission.}
{The remnant of T Pyx can be described by a two-component model, consisting of a tilted ring at $i = 63.7$ relative to its normal vector and by fast bipolar ejecta perpendicular to the plane of the equatorial ring.  }
{We find an upper limit for the bipolar outflow ejected mass in 2011 of the bipolar outflow of $M_{ej,b} < (3.0 \pm 1.0) \times 10^{-6}$ M$_{\odot}$, which is lower than previous estimates given in the literature. However, only a detailed physical study of the equatorial component could provide an accurate estimate of the total ejecta of the last outburst, a fundamental step to understand if T Pyx will end its life as a type-Ia supernova.}


\maketitle

\section{Introduction}\label{sec:1}

Classical Novae (CNe) are thermonuclear explosions that occur in binary-star systems consisting of a white dwarf (WD) accreting matter from a main sequence or evolved companion (for recent reviews see {\citet{Bode2008}, \citet{DellaValle2020}). Recurrent Novae (RNe) are CNe with more than one observed outburst since mankind has started to observe and report the occurrence of unexpected phenomena systematically. This class of stellar explosions holds significant interest as it stands as one of the most plausible astrophysical candidates for being progenitors of type-Ia supernova explosions, within the single-degenerate scenario \cite{Whelan1973,Nomoto1984}. It is worth noting that while it may not be the primary channel of SNe-Ia formation \citep{DellaValle1996,Shafter2015}, its significance remains prominent.

T Pyx is one of the most interesting RNe known to date, with six recorded outbursts observed so far, with the last outburst observed in 2011 \citep{Shore2011,Chesneau2011,Chomiuk2014,Joshi2014,Nelson2014,Pavana2019}. However, despite being one of the best-studied objects in this field, a complete understanding of the nature of the extremely massive primary white dwarf is still far from being achieved. It is unique, as T Pyx has the shortest orbital period among RNe \citep{Bode2008}, and 
it displayed a slow light curve evolution and a low magnitude amplitude compared to quiescence \citep{Surina2014}.
The rate of mass transfer inferred from historical UV observations at quiescence \citep{Gilmozzi2007}, and the slow luminosity decay during the outburst are in disagreement with theoretical expectations: the WD progenitor of RNe is expected to have a large mass to ignite the thermonuclear runaway (TNR, \citet{Gallagher1978,Starrfield2016}) on a short timescale, thus implying light ejecta and then a very fast luminosity decay during the outburst \citep{DellaValle2020}. Consequently, for a mass of the WD progenitor of $M \sim 1.35$M$_{\odot}$ \citep{Starrfield1985,Webbink1987,Schaefer2010}, and an accretion rate, obtained using several methods, of $\dot{M} = 1.1 \times 10^{-8}$ M$_{\odot}$ yr$^{-1}$, \citet{Selvelli2008} concluded that the mass ejected in each outburst of T Pyx is larger than the mass accreted during the quiescent period between two consecutive outbursts. This suggests that the primary WD will never reach the Chandrasekhar limit, 
and explode as a type Ia supernova (SN) \citet{Selvelli2008} \citep{Patterson2017}. On the other hand, \citet{Uthas2010} inferred a much lower value for the WD mass ($M \sim 0.7$M$_{\odot}$) and a mass ratio $q = M_2 / M_1 = 0.2$, where the subscript (1) stands for the primary WD \citep{Warner1973}. 
This finding would raise a theoretical problem for the observed short recurrence time. 

On the other hand, using a standard accretion disk model, and UV observations of T Pyx obtained with HST a few years after the last outburst, in addition to archival IUE and GALEX data, \citet{Godon2014} obtained a higher accretion rate of $\dot{M} = 10^{-6}$ M$_{\odot}$ yr$^{-1}$ for a distance of 4.8 $\pm$ 0.5 kpc \citep{Sokoloski2013}, which implies that the matter accreted onto the WD, independently of its mass, is larger than the gas expelled in the last 2011 outburst. Interestingly, the recent distance measurement obtained by the GAIA spacecraft \citep{GAIA2021} gives a value of $\sim$ 3.7 kpc (a value that has been estimated using the \citealt{BailerJones2018} likelihood model, and the volume density prior given in \citealt{BailerJones2015}).

The remnant of T Pyx has been often observed in the last decades, both with ground telescopes \citep{Duerbeck1979,Williams1982,Shara1989} and with the Hubble Space Telescope \citep{Shara1997,Sokoloski2013,Shara2015}. HST observations made with the F657N filter, covering the H$\alpha$+[N II] wavelength region, revealed the presence of thousands of knots, which are slowly and radially expanding away from the central WD \citep{Shara2015}. Using the light echo scattered by surrounding dust, \citet{Sokoloski2013} reported that 1) the remnant is mainly concentrated around a ring with an inclination of 30-40 degrees, and 2) the distance to the remnant is $d = 4.8 \pm 0.5$ kpc. The most recent observations of T Pyx with HST, executed three years after the last 2011 outburst, show also the presence of a two-lobed structure, oriented along the East-West direction, and that were absent in the HST observations made before the 2011 outburst. The left panel of Fig. \ref{fig:3} shows the central region of the remnant observed with the F657N narrow filter, while in the right panel, we have reconstructed the emission in the same wavelength range covered by the F657N filter using MUSE observations. This dataset will be discussed later in Sec. \ref{sec:4}.   

High-resolution spectroscopic observations several years post-explosion show also evidence for different types of aspherical ejecta \citep{Chesneau2011,Shore2011,Shore2013,Aquino2014,Izzo2015b}. Using HST-UV echelle observations of the nebular phase following the last outburst, \citet{Shore2013} explained the profiles observed in the main nebular emission lines with an axisymmetric bipolar geometry. However, our ignorance of the exact geometrical distribution of nova ejecta has an immediate effect on the estimate of the mass ejected in a nova outburst.  
The difference in the ejected mass inferred from the above models varies by one order of magnitude. This seems in agreement with the recent scenario proposed by \citet{Mukai2019,Aydi2020b} where nova outbursts are characterized by two main ejection flows, each with different velocities, geometrical distribution, and origin, which interact giving rise to shocks during the brightest nova emission.  

\begin{figure*}
    \centering
    \includegraphics[width=0.48\linewidth]{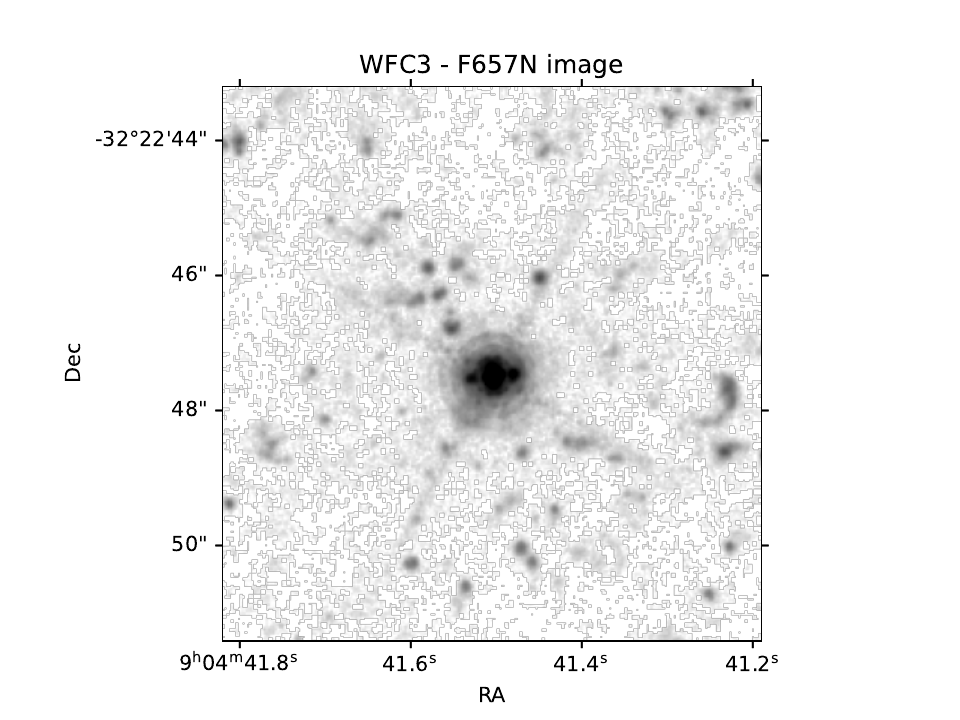}
    \includegraphics[width=0.49\linewidth]{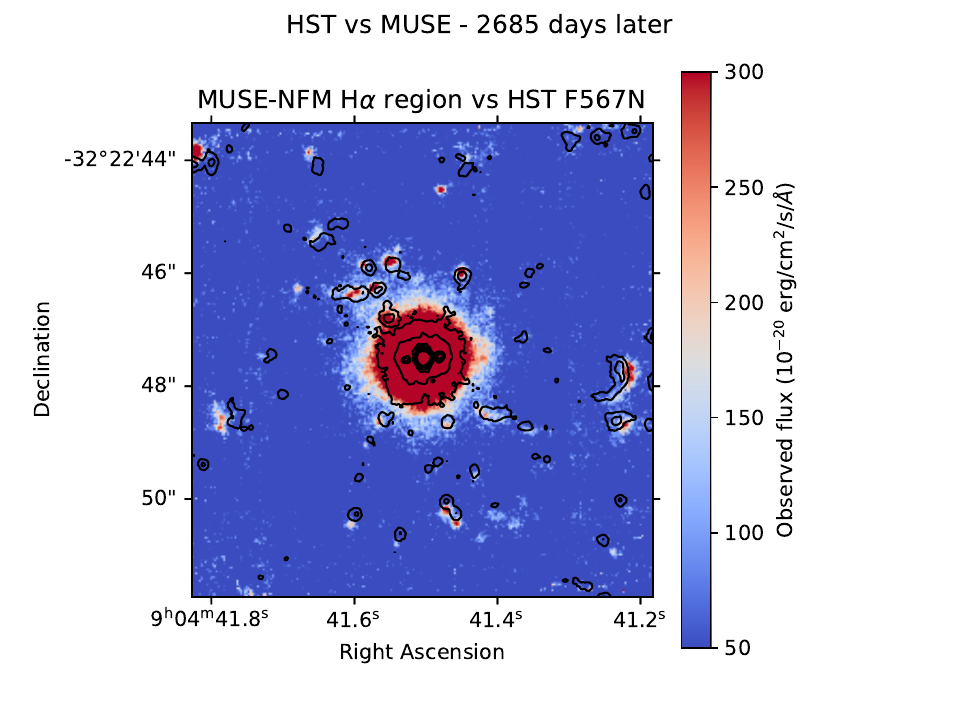}
    \caption{(Left panel) The HST/WFC3 image of the central region, with the same size as the MUSE-NFM of the T Pyx remnant, obtained three years after the last outburst, and with the F657N filter, which is centered at the H$\alpha$ emission line. 
    (Right panel) The H$\alpha$ flux map, reconstructed from the MUSE-NFM datacube. Black curves correspond to iso-surface brightness contours derived from the HST/WFC3 F657N image. The image scale is the same, but MUSE data have been obtained 7.35 years (2685 days) after HST data. This reflects on the clear shift for the majority of the H$\alpha$-[N II] knots visible in the image. }
    \label{fig:3}
\end{figure*}

To shed light on the physical properties of the ejecta of T Pyx, and on its possible role as one of the potential type-Ia SN progenitors in our Galaxy, we have observed the remnant with new eyes, namely with the Multi Unit Spectroscopic Explorer (MUSE) mounted at the UT4 (Yepun) of the European Southern Observatory (ESO) at the Cerro Paranal, Chile. In this work, we report on the analysis of the geometrical distribution of the remnant, and the estimate of the distance to T Pyx using the nebular parallax method applied to the expanding ejecta from the last outburst in 2011. In Sec. \ref{sec:2} we initially describe the observations executed with MUSE, while in Sec. \ref{sec:3} we report on the analysis of the datacube obtained using the Wide Field Mode (WFM), and then on the extended remnant. In Sec. \ref{sec:4} we move to the analysis of the Narrow Field Mode (NFM), which allowed us to obtain a direct distance estimate to the system. In Sec. \ref{sec:5} we discuss our findings and in Sec. \ref{sec:6} we draw our conclusions. 

\section{Observations}\label{sec:2}

The remnant of the RN T Pyx has been observed with MUSE \citep{Bacon2010}, mounted at the UT4 (Yepun) at the European Southern Observatory / Very Large Telescope, located in Paranal, Chile. MUSE is an integral field unit spectrograph, capable of providing spatially-resolved spectroscopy in the visible range. Two modes are available with MUSE: 1) a Wide Field Mode (WFM), which allows to cover a region-wide 1$\times$1 arcmin, with a nominal spatial sampling of 0.2 arcsec, and a spectral resolution variable from R $\sim$ 1800 at 4800 \AA\, and R $\sim$ 3600 at 9300 \AA\, of 2) a Narrow Field Mode, which allow to observed with a higher spatial resolution (sampling of 0.02 arcsec) a narrower region of 7.5$\times$7.5 arcsec, with a similar spectral resolution. With the WFM, the ground layer adaptive optics (GLAO) correction is used, which provides a correction of 0.2$\times$0.2 arcsec over the entire field of view, while in NFM the laser tomography adaptive optics (LTAO) is used that provides a high correction, with a diffraction core and a Strehl ratio between 5$\%$ and 10$\%$ at H$\alpha$. The use of AO, however, implies a gap in wavelengths of about $\sim$ 270 \AA\, around the \ion{Na}{I} D region (5890 \AA) due to the use of the four sodium laser guide-stars. 

  The NFM spatial resolution is comparable to that of HST images, but in addition to this, we also have the spectral information from each spaxel, which provides us with the radial velocity distribution from emission lines. This information can be used as a proxy for the "third" spatial dimension to reconstruct the 3D-geometric distribution of the explosion and obtain an accurate estimate of the distance. NFM observations were done on November 15th, 2021 (corresponding to 3868 days after the last outburst, and to 2685 days, e.g. 7.35 years, after the HST observations reported in Fig. \ref{fig:3}, and that are considered as our reference for the nebular parallax analysis in Sec. \ref{sec:4}), while WFM observations have been executed on January 1st and 9th, 2022 (see also Table \ref{tab:1}. Data have been reduced using standard \texttt{esorex} recipes embedded in a single Python-based pipeline. After the creation of static calibration datacubes, namely bias, flat, and wavelength calibration, we have also built the line-spread function and the illumination correction. Then, we applied the above corrections to the science and the standard datacubes and built the corresponding pixel tables for each single sensor. Finally, we have obtained the response function for the standard star and obtained the final stacked science datacube. Astrometry has been computed using HST archival data, with the central WD as the main reference. Sky emission residuals have been subtracted using \texttt{Zap} \citep{Soto2016} after masking all the sources in the field of view of MUSE-WFM, whose emission is lower than the detection threshold of 1.5 sigma. 

\begin{table}
\small
\centering
\caption{Diary of MUSE spectral observations}
\label{tab:1}
\begin{tabular}{lcccc}
\hline \hline
Date & MJD & Range & Mode & Days after HST \\
(days) & (days) & (\AA) & & \\
\hline
Nov 15, 2021 & 59533.31 & 4750-9350 & NFM & 2685 \\
Jan 1, 2022 & 59580.21 & 4600-9350 & WFM & 2732 \\
Jan 9, 2022 & 59588.16 & 4600-9350 & WFM & 2741 \\
    \hline
    \hline
        \end{tabular}
\end{table}

\section{The extended remnant}\label{sec:3}

One of the great advantages of using IFU data consists in the possibility to reconstruct 2D images centered on a specific wavelength range, and then in the case of nova remnants to focus on the emission from a given line feature. This allows us to recover, for example, the emission from the hydrogen component alone, after subtracting the integrated flux in a given wavelength window region for the underlying stellar continuum of the central WD. In this way, 2D images of the expanding hydrogen ejecta have been built from the WFM dataset by centering around the H$\alpha$ ($\lambda = 6562.8$ \AA) and H$\beta$ ($\lambda = 4861.3$ \AA) emission line, with an extraction wavelength window of $\Delta \lambda = 20$ \AA\,.  The final results show the presence of a tilted clumpy ring structure with an apparent semi-major radius of $\approx$ 5-6 arcsec, see Fig. \ref{fig:1}. To model this component, we use the observed radial velocity as a proxy for the third spatial dimension $z = v_\mathrm{rad} t$, given that at this epoch the ejecta is completely frozen and freely expanding. Using these flux maps, and the distribution of the radial velocity of the gas along the hydrogen ring ejecta, we can determine the properties of the remnant such as the position angle $PA$ and the inclination angle of the ring.

However, due to the velocity distribution of the remnant, we could not estimate the de-projected radial velocity of the expanding gas using H$\alpha$, because of the significant contamination from the nearby [N II] 6548/84 \AA\, doublet. On the other hand, when focusing on the H$\beta$ line, which is quite isolated from other contiguous bright emission lines such as [O III] 4959/5007 \AA\, from the remnant, and He II 4686 \AA\ from the central WD, we were able to isolate the H$\beta$ emission line, and finally fit it with a Gaussian function. In this way, we can reconstruct the velocity field of the expanding ring structure without any contamination from other emission lines, see the right panel in Fig. \ref{fig:2}. To perform this operation, however, we have first re-binned the entire WFM datacube using a Voronoi tessellation algorithm, previously masked for external field sources. This technique provides a spatially-binned datacube based on the value of the signal-to-noise ratio in a specific wavelength region \cite{Cappellari2003}. Since the H$\beta$ line is fainter than H$\alpha$, we use the wavelength region centered on the H$\alpha$ line with a width of $\delta \lambda$ = 12.5 $\AA$, and a signal-to-noise threshold of 10. After this, we calculated the radial velocity of each spatial bin by fitting the H$\beta$ emission line using a Gaussian function, to measure the corresponding velocity offset, after correcting for the barycentric velocity of T Pyx at the epoch of observations ($v_\mathrm{b} = 16.5$ km/s). The inner ($r \leq 5''$) region has been masked while computing the radial velocity offset, since it is dominated by the light from the central WD. The final H$\beta$ radial velocity map of the remnant region outlined by the over-plotted contours from the H$\beta$ flux map presented in Fig. \ref{fig:1}, is shown in Fig. \ref{fig:2}.

\begin{figure*}
    \centering
    \includegraphics[width=0.48\linewidth]{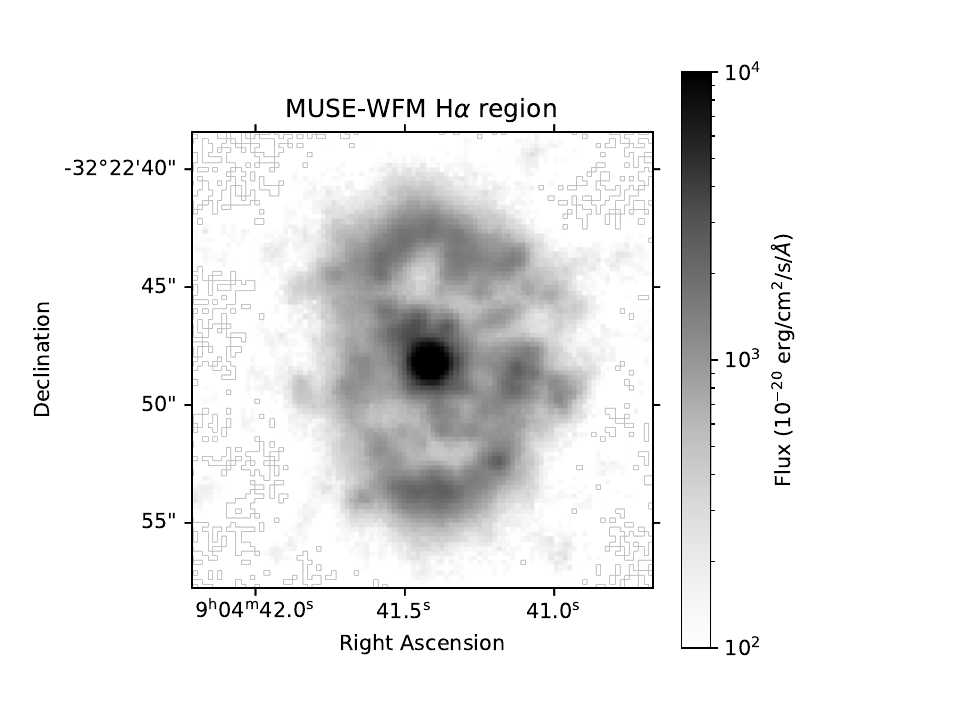}
    \includegraphics[width=0.48\linewidth]{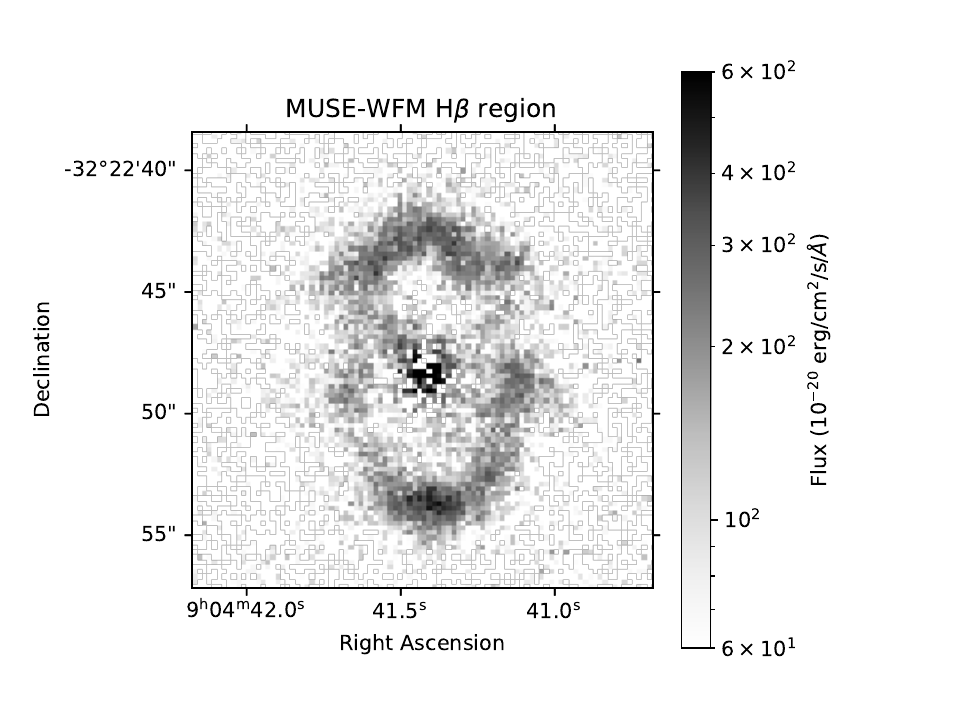}
    \caption{The reconstructed H$\alpha$ image of the remnant (left panel) and the corresponding H$\beta$ image (right panel), both obtained from the MUSE-WFM data as described in the main text. }
    \label{fig:1}
\end{figure*}

To recover the geometrical properties of the 3D ring remnant from the observed 2D map of Fig. \ref{fig:2}, we initially simulated the three-dimensional geometry of the entire ring remnant for different values of the ring radius, expanding velocity, inclination, and position angle. To build our ring model, we have used a Cartesian representation embedded within a \texttt{python} script developed for this analysis\footnote{https://github.com/lucagrb/TPyx}. The initial 2D ring equation, which does not yet include any inclination along the observer's line of sight, can be written as follows:

\begin{figure*}
    \centering
    \includegraphics[width=0.48\linewidth]{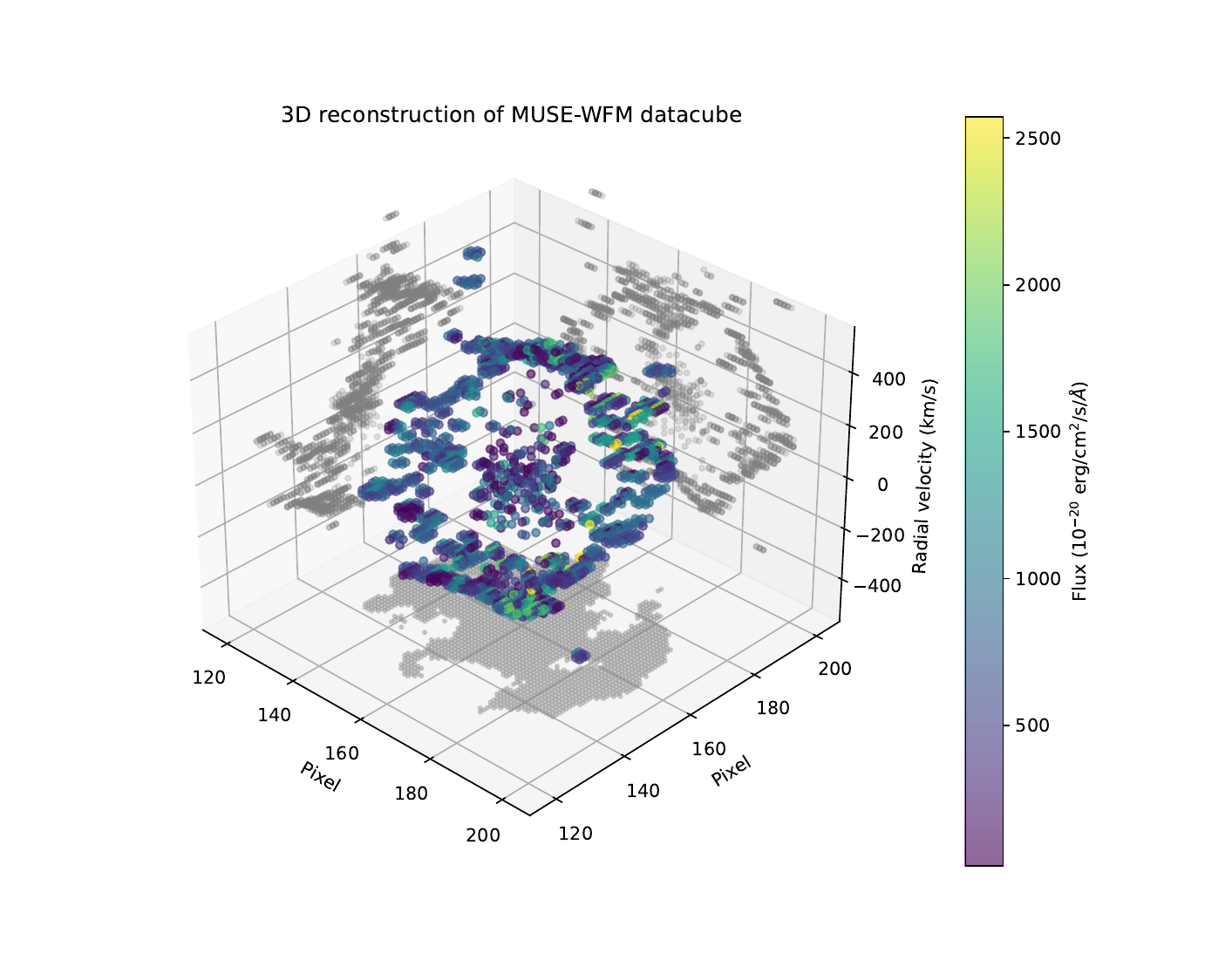}
    \includegraphics[width=0.48\linewidth]{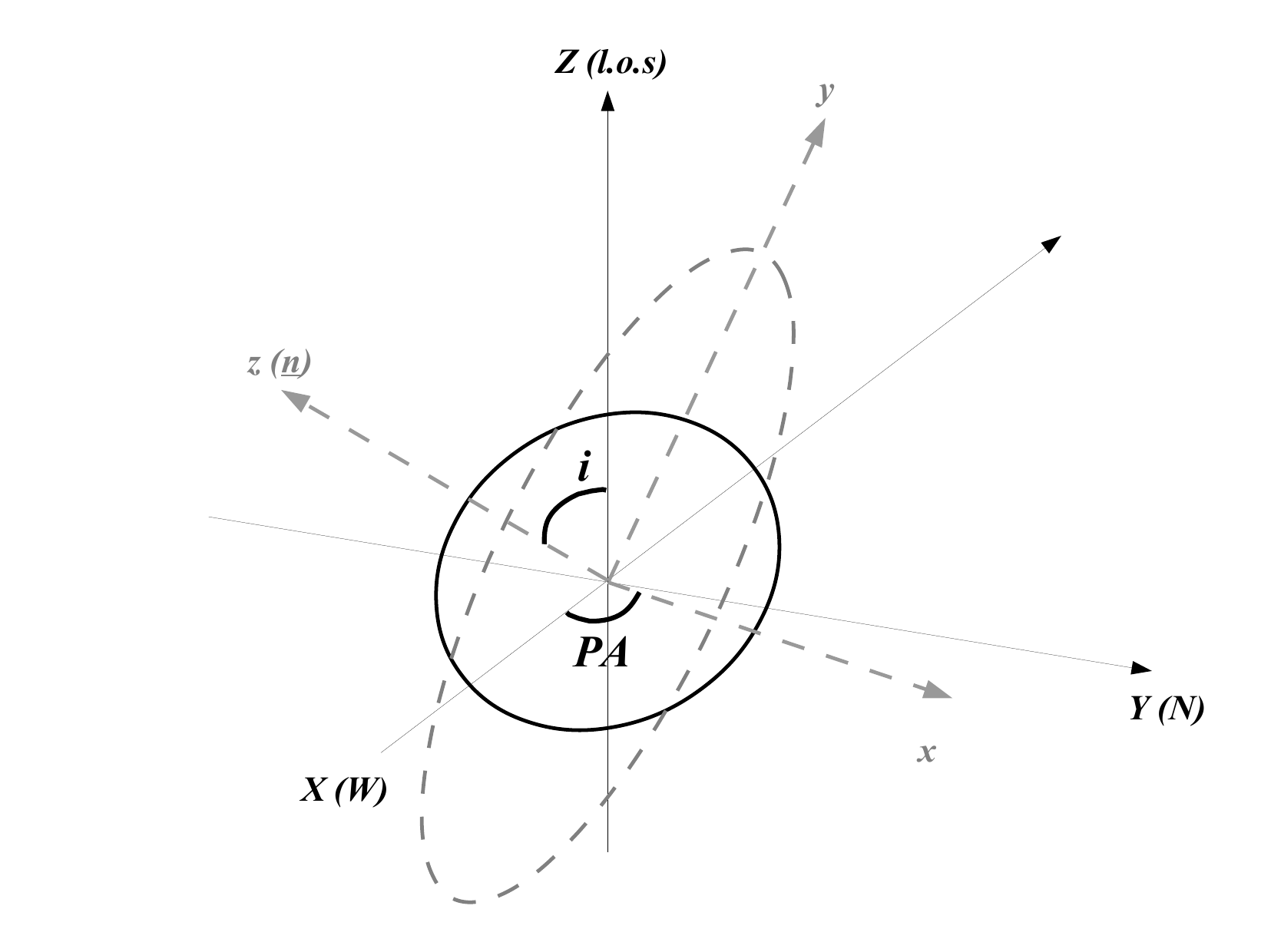}
    \caption{(Left panel) The 3D reconstruction of the T Pyx remnant using the spatial resolution provided by the MUSE-WFM and using the radial velocity from H$\beta$ emission line as a proxy for the third dimension. The projection onto the corresponding 2D planes is shown with gray data. Each single data point corresponds to a single bin, resulting from the Voronoi tessellation algorithm described in Sec. \ref{sec:3}. It is visible how the external remnant is arranged as a non-uniform thick ring, with inhomogeneities marked by the different flux intensities of the line (reported here with different colors). The line of sight direction is perpendicular to the X-Y (pixel-pixel) plane, and here it is shown as pointing to the top direction. (Right panel) The reference frame used to model the remnant of T Pyx, and to recover the transformation equations described in Eq. \ref{eq:3}, is shown in black, with the observer's line of sight located along the Z axis. The intrinsic remnant reference frame is reported with gray coordinates and frame. The inclination ($i$) and position angle ($PA$) are also shown to help to reconstruct the projected model.}
    \label{fig:1}
\end{figure*}

\begin{equation}\label{eq:1}
\begin{array}{l}
    x = (R + r \times \cos{\phi}) \times \cos{\theta}\\
    y = (R + r \times \sin{\phi}) \times \sin{\theta}
\end{array}
\end{equation}

where $\phi$ and $\theta$ are two angles modeling the ring and the width of its border, and go over the (0, 360) deg range, $R$ is the width of the ring, and $r$ (fixed to 1 pixel) is the thickness of the ring border. When including tilt effects on our simulated ring, we have taken into account two other important parameters: 1) the inclination angle $i$, corresponding to the angle between the normal vector to the ring and the line of sight, and 2) the position angle $PA$, which takes into account any (anti-clock-wise) rotation of the tilted ring between the East direction in the sky and the ring radius connecting the ring center to the position of observed "zero radial velocity". This representation is very similar to the tilted disk model used to study galaxies and AGN \citep{Begeman1989}, with the only difference that in this case, the gas is expanding radially, while for galaxies the gas is rotating.

Due to the symmetry of the model, there are two ring radii connecting the center to the zero velocity value, with the two radius lying on the same direction: to exclude any degeneracy, we fix a prior on the PA value that can vary between 0 and 180 degrees. The inclusion of these two parameters provides us with the projected 2D equations, including the observed radial velocity value parameterized by the $Z$ component:

\begin{equation}\label{eq:1}
\begin{array}{l}
    X = x \times \cos{PA} - y \times \sin{PA} \times \cos{i}\\
    Y = x \times \sin{PA} + y \times \cos{PA} \times \cos{i}\\
    Z = y \times \sin{i} \times v'
\end{array}
\end{equation}

Here, to recover the intrinsic expanding velocity, we should correct $v'$ for the projected $Z$ component, namely $v_\mathrm{exp} = v' * R / \sin{i}$. A graphical representation of our model is shown in Fig. \ref{fig:1}, where the tilted ring parameters correspond to the best-fit values (see below).

Now, we have to compare our model with the observed 2D ring radial velocity distribution by projecting the resulting 3D ring geometry distribution onto the 2D spatial plane, corresponding to the observed image of the T Pyx remnant. The expanding velocity is assumed constant along the entire 3D distribution of the ring, which is a valid assumption for the frozen expanding gas at $\sim$ 10 years after the last outburst.  Values of radial velocities have been simulated for 180 data points along each projected ring (this has been obtained through sampling the $\phi$ and $\theta$ angle parameters with a fixed separation between each pair of data of 2 degrees), and finally fitted to the observed radial velocity distribution. For this last operation, we have performed the fit using a Markov-Chain Monte-Carlo (MCMC) method using the \texttt{emcee} package \citep{Foreman2013}. This MCMC methodology estimates the values of our unknown parameters by constructing a posterior distribution for all of them, through the use of a likelihood function and a set of uniform priors for each of the parameters, which have been at first determined through an initial grid-search fit. Our Bayesian likelihood function is the classical chi-square function, with uncertainty values given by the observed radial velocity dispersion (obtained from the corresponding error datacube) for the relevant spatial bin of each observed data point. 

The results of our MCMC fitting procedure are summarized in Fig. \ref{fig:R}. We find that the remnant of T Pyx is best modeled by a ring structure with a radius of $R = 5.6^{+0.7}_{-0.7}$ arcsec, an expanding velocity of $v_\mathrm{exp} = 471.7^{+77.2}_{-72.2}$ km/s, an inclination angle of $i = 63.7^{+20.8}_{-13.8}$ deg, and position angle of $PA = 106.7^{+39.2}_{-40.3}$ deg. Our results are in good agreement with results obtained from the light echo analysis by \citet{Sokoloski2013}. Moreover, we find a very good agreement with the position angle found using interferometric observations by \citet{Chesneau2011,Pavana2019}, but considering, in their analysis, a bipolar ejecta geometry. Assuming a GAIA distance of 3.7 kpc (see above), and the expanding velocity determined above, we obtain an extension of the ring of $R_\mathrm{phys} = (2.76 \pm 0.33) \times 10^{12}$ km, which would have been ejected $t_\mathrm{ej} = 207 \pm 35$ years ago by the central WD. This value would correspond to an eruption that happened in a relatively large time interval between 1779 and 1849. This result marginally agrees with the estimate by Schaefer et al. of a CN-like outburst of T Pyx in 1866 \citep{Schaefer2010}, an estimate that however considered for T Pyx a slightly lower distance of 3.5 kpc. A larger distance of $d = 4.8 \pm 0.5$ kpc, as recently proposed by \citet{Shore2013,Sokoloski2013} would only agree within 2-$\sigma$ with \citet{Schaefer2010} estimating, resulting in a CN-like ejection epoch that happened of $t_\mathrm{ej} = 270 \pm 56$ years ago. T Pyx should be at a closer distance to be in full agreement with the \citet{Schaefer2010} calculation. An accurate distance of T Pyx can shed light on the origin of the ring structure. We remark that we have not considered any deceleration of the ejecta due to the snowplow effect by interaction with the surrounding CSM or with gas from previous eruptions. An extended remnant initially expanding at larger velocities would suggest a more recent CN-like eruption, and then agree with \citet{Schaefer2010}. A more detailed analysis of the extended remnant will be presented in a further publication.

\begin{figure*}
    \centering
    \includegraphics[width=0.49\linewidth]{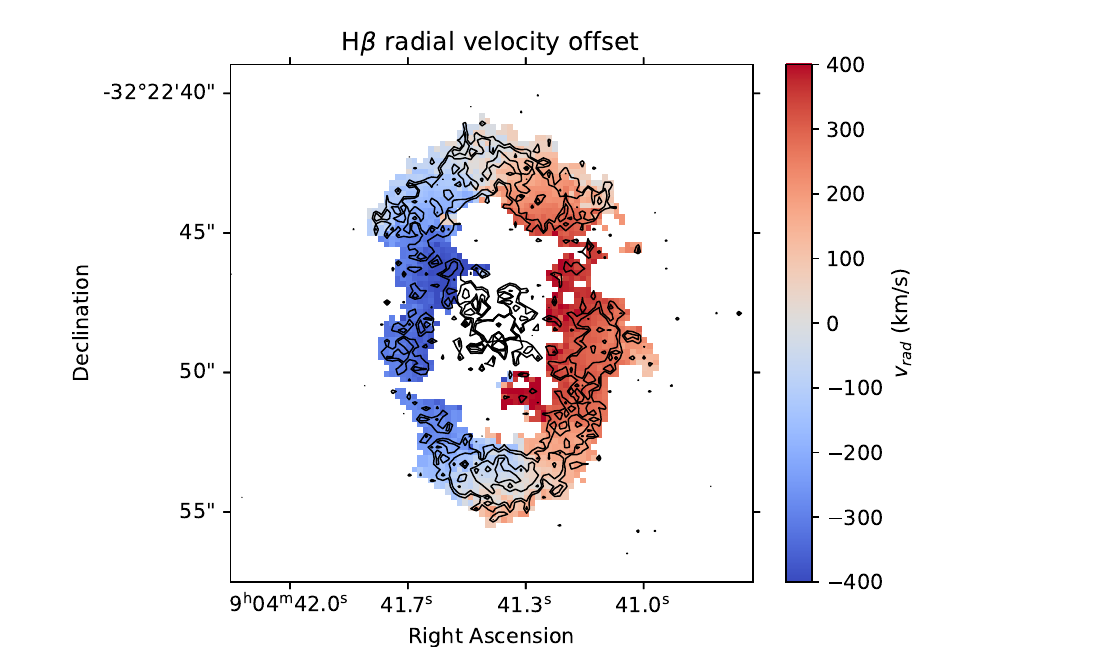}
    \includegraphics[width=0.46\linewidth]{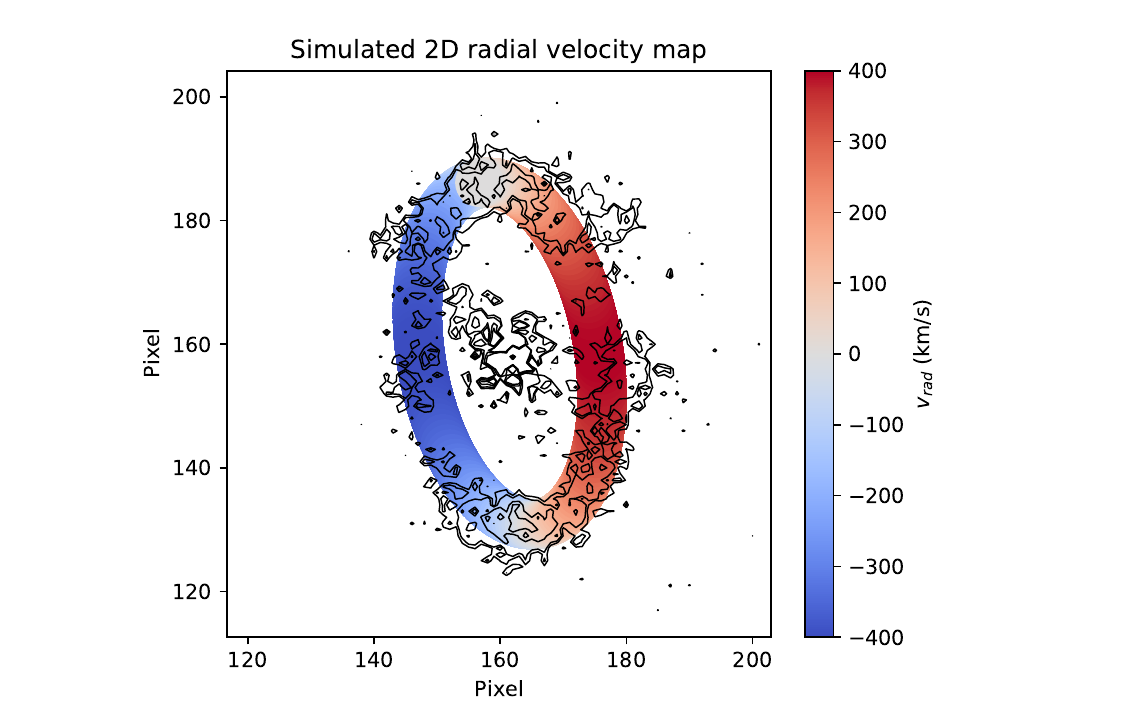}
    \caption{(Left panel) The H$\beta$ radial velocity map, obtained with the prescriptions given in the main text. The black curve represents the contour regions from the H$\beta$ intensity flux map of Fig. \ref{fig:1}. (Right panel) The simulated 2D radial velocity map of the ring remnant was obtained from the best-fit parameters and the procedure delineated in the main text.}
    \label{fig:2}
\end{figure*}

\begin{figure}
    \centering
\includegraphics[width=0.98\linewidth,height=220pt]{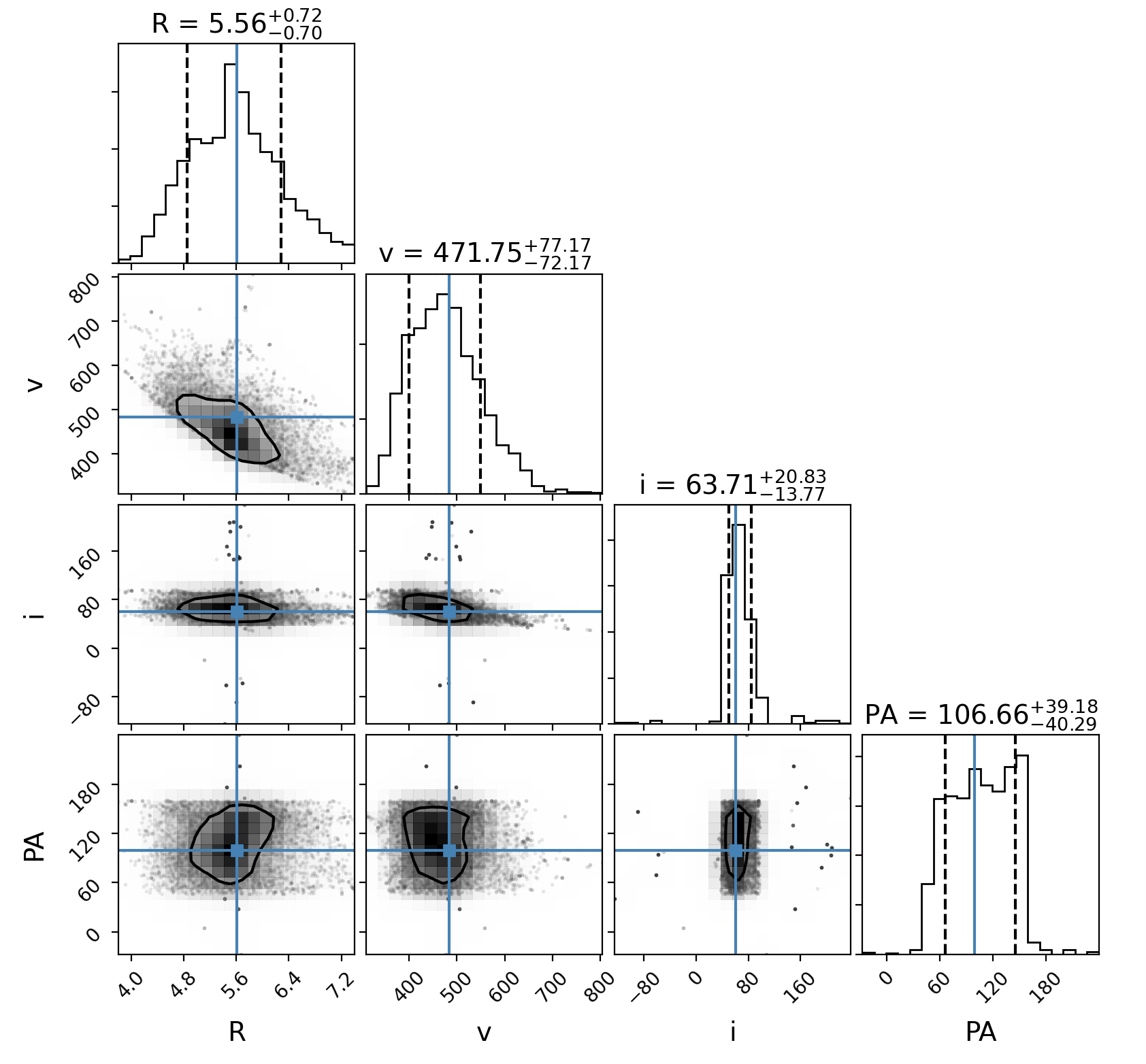}
    \caption{the posterior distribution and corner plots obtained from the MCMC fitting procedure described in Sec. \ref{sec:3}.}
    \label{fig:R}
\end{figure}

\section{The 2011 outburst ejecta}\label{sec:4}

The possibility of combining the spectral information of spatially identified structures allows us to obtain an accurate measurement of the distance to T Pyx using the nebular parallax method \citep{Baade1940,Cohen1985,Krautter2002}. The remnant of T Pyx has been observed a few times with HST after the last outburst of 2011 \citep{Toraskar2013,Godon2014,Shara2015}. In one of the most recent observations, a bipolar ejecta is clearly identified along the WE direction, in particular using the narrow filters F657N available with the Wide Field Camera (HST GO program ID: 13796, PI: Crotts), see also Fig. \ref{fig:3}. We have searched for the presence of evidence of this bipolar structure in the MUSE-NFM datacube, which has a very high spatial resolution of $\sim$ 0.03 arcsec, comparable to the HST/WFC3 resolution. First, we studied the reconstructed H$\alpha$ map and detected all the knots observed by HST in 2014. For the majority of them, we also notice a shift in the spatial coordinates due to their expanding motion. We also identified an extended knot in the W direction with respect to the central WD, which was absent in pre-2011 outburst HST images, and that is confirmed to be the expanding outflow of the 2011 outburst only thanks to the reconstruction of the [O III] map. Indeed, late nebular spectra of the 2011 outburst of T Pyx have shown the presence of bright [O III] lines a few years after the initial explosion of the nova \citep{Shore2013,Pavana2019}. Unlike the H$\alpha$ line, the wavelength region surrounding the [O III] $\lambda\lambda$ 4959/5007 doublet is far from bright lines observed in emission from the central WD and its accretion disk (He I 4922 and 5016 $\AA$ lines are visible only in the innermost regions where light from the central WD dominates). The [O III] emission line map covering a region of $\sim$ 2 arcmin radius from the central WD is shown in Fig. \ref{fig:4}.

\begin{figure*}
    \centering
    \includegraphics[width=0.48\linewidth]{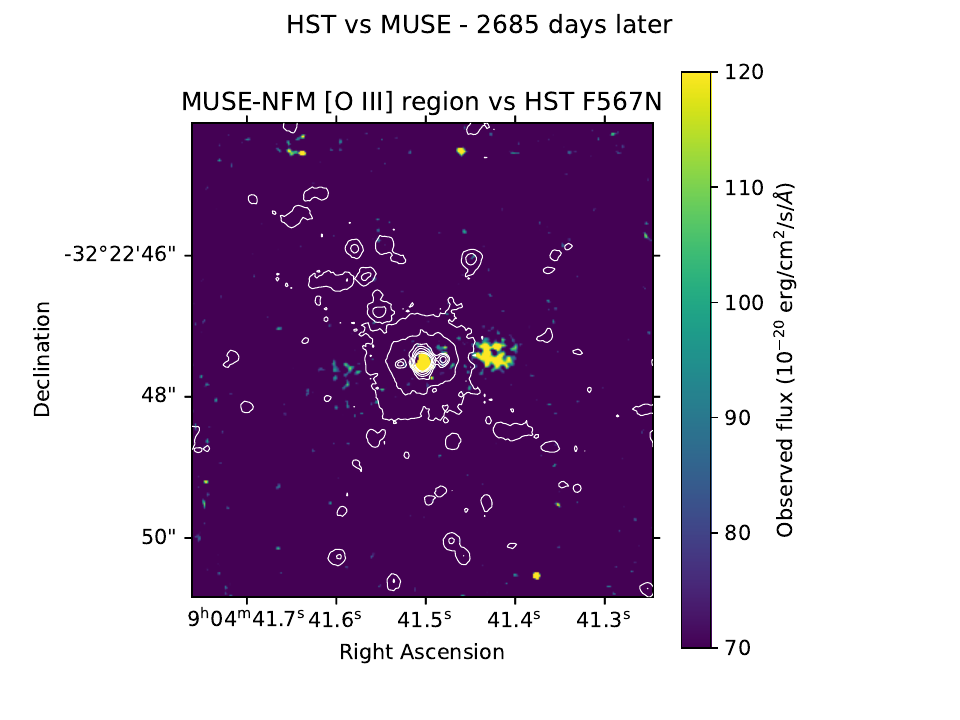}
    \includegraphics[width=0.49\linewidth]{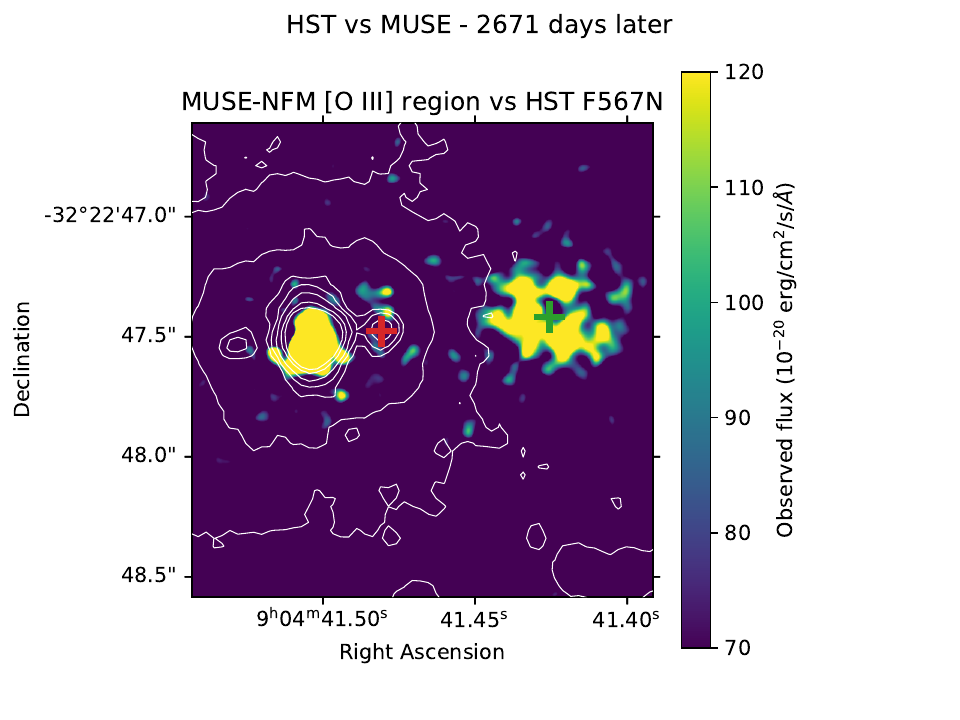}
    \caption{(Left panel)  (Right panel) A 2'$\times$2' zoom into the region covering the central WD and the bright pole ejecta. After registering the MUSE [O III] map for the HST astrometry, we measure a shift of  0.71 arcsec between the centroid positions of the ejecta observed in 2014 (with HST - green cross) and in 2021 (with MUSE - red cross). }
    \label{fig:4}
\end{figure*}

The resulting [O III] map clearly shows the presence of two extended structures at larger, and opposite, distances from the central WD if compared to the HST observation made 7.35 years before the MUSE observations. The western lobe turns out to be brighter than the eastern one, and it also shows some sub-structures, suggesting a complex shape for the ejected mass. The integrated spectra of both blobs, extracted from the MUSE-NFM dataset, and subtracted for underlying residual WD emission using a background region with the same pixel area as the source extraction region, shows that the western lobe is blue-shifted ($v_{rad} = -798 \pm 50$ km/s) with respect the eastern counterpart (see Fig. \ref{fig:5}).  Interestingly, the H$\beta$ line shows only the component from the underlying WD and is located at the systemic velocity of T Pyx. A faint residual emission is visible at bluer wavelengths with the same blue shift reported for the western blob. While this evidence explains the fainter emission from the eastern lobe, being more distant and then passing through a thicker circumstellar medium, it also confirms that the bipolar ejecta are directed toward the perpendicular direction to the plane of the larger ring-like structure, characterized previously using the MUSE-WFM dataset (see Sec. \ref{sec:3}). This characteristic allows us to use the information obtained for the spatial orientation of the ring-like structure to determine the velocity component of the bipolar ejecta tangential to the line of sight. Then, we can use the yearly parallax of the bipolar ejecta, after determining the separation between the western blob observed in the HST image of 2014 and the MUSE/NFM 2021 datacube, with the information on the tangential velocity to determine the distance using the nebular parallax method. 

\begin{figure}
    \centering
    \includegraphics[width=0.88\linewidth]{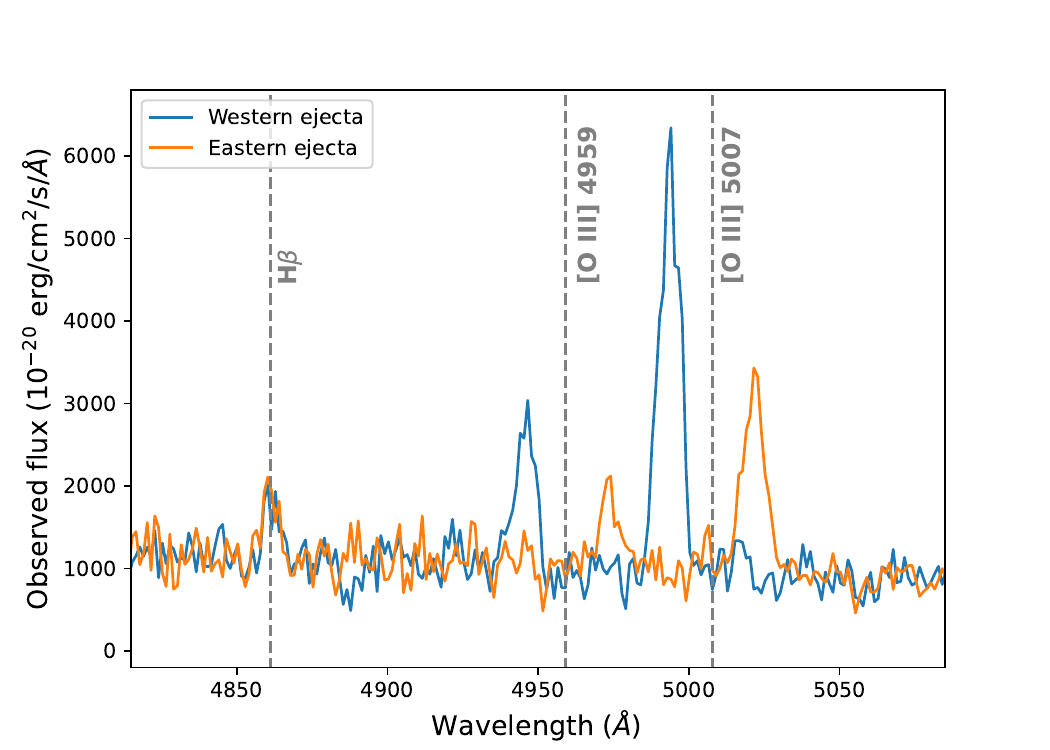}
    \caption{The wavelength region between 4800 and 5100 \AA\, of the western blob ejecta (blue data) and the eastern blob (orange data) detected in the MUSE-NFM dataset. Dashed gray lines mark the rest-frame wavelength for the H$\beta$ and the [O III] 4959/5007 lines. While H$\beta$ is mainly detected at zero velocity offset, suggesting it mainly comes from the gas surrounding the progenitor WD, [O III] lines show a radial velocity offset originating from the ejecta of the 2011 outburst.}
    \label{fig:5}
\end{figure}

A more detailed insight on the [O III] map also shows the presence of a structure in the western blob ejecta: while the HST image shows a compact spheroidal knot, NFM data reveal the presence of two elongated filaments that spread laterally to the main compact ejecta. We notice that a similar structure was already inferred in 1D spectral analyses \citep{Shore2013,Pavana2019}. To determine the position of the western blob in both datasets, we have first determined the centroids of the blue ejecta from the HST/WFC3 image and the MUSE-NFM [O III] map, after being corrected for the HST astrometry. For the exact measurement of the centroids, we have used the \texttt{centroid\_com} package, available within the \texttt{photutils} python package \citep{Bradley2022}. In particular, we have first selected the region of the HST and MUSE-NFM [O III] maps (after being registered for the astrometry of the HST-WFC3 image) containing the western blob, and then we have calculated the centroids of the corresponding observed light distribution by estimating the center of mass determined from moments of the light distribution. We have finally reconverted the centroid values into World Coordinate System (WCS) coordinates using the astrometry of the HST image. The centroids of the western blob in the two epochs are reported in Table \ref{tab:2}, while the results of our analysis are shown in Fig. \ref{fig:4}. The resulting distance of the two centroids amounts to $d = 0.69 \pm 0.04 (stat) \pm 0.05 (sys)$ arcsec. 

\begin{table}
\small
\centering
\caption{Centroid positions (in image and WCS J2000.0 coordinates) of the western blobs in the HST and MUSE-NFM maps}
\label{tab:2}
\begin{tabular}{lcccc}
\hline \hline
Epoch & X & Y & RA & Dec. \\
 & (px) & (px) & (deg) & (deg.)\\
\hline
2014.52 & 779.9752 & 795.2022 & 136.172838 & -32.379854 \\
2021.87 & 798.2706 & 799.3953 & 136.172639 & -32.379815 \\
    \hline
    \hline
        \end{tabular}
\end{table}

The computed centroids of the western blob at both epochs display a tilt of $\phi = 5.3$ deg with respect to the WE direction. Assuming that the direction of this ejecta component is normal to the plane of the ring distribution derived in Sec. \ref{sec:3}, this value is in agreement within 1 sigma with the PA value found in the previous analysis, see Sec. \ref{sec:3}. This hypothesis is also supported by the evidence that the western blob displays a blue-shifted, and larger, radial velocity while the western ring component shows red-shifted and lower, velocity values. This evidence also supports the proposal of a multi-component ejection in T Pyx. However, we do not find clear evidence of a ring component from the 2011 outburst, as well as from previous outbursts. This can be attributed mainly to two factors: 1) a low spatial resolution of the available dataset, which prevents us from disentangling the ring from the nearby bright WD, and 2) the possibility that the ring component is not massive enough to emit hydrogen spectral lines detectable with MUSE. Another possibility, which however needs more investigation, is that the more recent T Pyx outbursts only displayed a bipolar ejecta, while the massive ring component was produced in a more energetic explosion that happened further back in time.  Indeed, detailed nebular analyses have excluded the presence of continuous wind in the last outburst, and the best model that fits the observed line structure is provided by an axial bipolar symmetry \citep{Shore2013,Aquino2014}. A complete physical characterization in terms of ejecta constitution and abundance of the extended remnant will be studied in a forthcoming publication.

\section{Discussions}\label{sec:5}

The characterization of the geometry distribution of the remnant of T Pyx paves the way to determine the distance to the nova using the nebular parallax method \citep{Baade1940,Cohen1985,Krautter2002}. We have determined the tangential offset displayed by the ejecta of 2011 as observed in two epochs separated by 7.35 years, but we still need information about the tangential velocity of the ejecta itself. This can be determined using the MUSE spectrum of the western blob, in particular through the measurement of the radial velocity from emission lines such as [O III], and then transform this into a tangential velocity value using the geometry of the remnant determined previously. The spectra of the bipolar ejecta both show a radial velocity offset of $v_{r,ej} = 798 \pm 50$ km/s (corrected for the measured systemic velocity of $v_{sys} = +28$ km/s), opposite in sign for the eastern blob, see also Fig. \ref{fig:5}. According to the assumed geometrical representation, see also Fig. \ref{fig:1}, we can determine the tangential velocity by correcting the observed radial velocity of the ejecta for the position ($PA$) and the inclination angle ($i$) through $v_{t,ej} = v_{r,ej} \times \tan{i}$. Using the results derived in Sec. \ref{sec:3}, we obtain a tangential velocity of $v_{t,ej} = 1613 \pm 340$ km/s. We have now all the ingredients to derive the nebular parallax distance, which is defined by:

\begin{equation}
    d = 0.207 \left( \frac{v_{t,ej}}{\dot{\theta}} \right) = 3.55 \pm 0.77 \,\textrm{kpc},
\end{equation}

where $\dot{\theta} = (9.4 \pm 0.5) \times 10^{-2}$ arcsec/yr, is the estimated expansion parallax, and $v_{t,ej}$ is given in km/s \citep{Wade1990,Duerbeck1992}. This distance value is in full agreement with previous distance estimates existing in literature for T Pyx \citep{Selvelli2008,Schaefer2010}, and with GAIA measurement \citep{GAIA2021}, but it is only marginally consistent with the value given in \citet{Shore2011,Sokoloski2013}.

With the knowledge of the distance to T Pyx, thanks to the geometrical reconstruction of the bipolar ejecta, and its size, it could be also possible to provide an estimate of the mass of the bipolar ejecta using the information available in the western blob spectrum alone, assuming that the same mass is expelled in both directions. A widely used methodology to infer nova ejecta masses from nebular phase spectra consists of measuring electron temperatures and accurate fluxes for Balmer lines (in this case the H$\beta$, being H$\alpha$ strongly blended with [N II]), so that one can directly determine the electron density in the ejecta and from the exact knowledge of the ejecta volume, the ejected mass \citep{Mustel1970}. Briefly, the luminosity of H$\beta$ line can be expressed by:

\begin{equation}\label{eq:3}
    L(H\beta) = \epsilon_{H\beta}\, n_e^2\, V
\end{equation}

where $\epsilon_{H\beta}$ is the emission coefficient for H$\beta$ at the temperature of $T = 10,000$ K, typical of case B recombination \citep{Osterbrock}, $n_e$ is the electron density, assumed to be equal to hydrogen density ($n_p = n_e$), and $V$ the volume of the ejecta. A drawback of this methodology consists in the lack of precise geometrical characterization of the ejecta itself, for which a general assumption of a spherical shell, corrected for the filling factor of the gas, is usually considered \citep[see e.g. ][]{DellaValle2003,Mason2005}. In the case of T Pyx, we have a direct measurement of the size of the western bipolar ejecta, assuming it is best characterized by a structured spherical blob with $r_{ej} = 0.22$ arcsec, which at the distance of $d = 3.55 \pm 0.77$ kpc results in a volume of $V = (5.43 \pm 0.90) \times 10^{48}$ cm$^3$. However, we cannot exclude the clumpiness of the ejecta \citep{Santamaria2022}, as can be deduced from the [O III] map in Fig. \ref{fig:4}, which would still give rise to a multiplicative filling factor term. Here we assume a conservative value of $f \eqsim 0.1$ for the filling factor, whereas \citet{Shore2013} found $f = 0.03$ (but using integrated 1D spectra).

The spectrum extracted from the western blob does not show any clear evidence for H$\beta$ at the same radial velocity of [O III] lines, which is the main signature of hydrogen emission from the ejecta. Assuming the same FWZI reported for [O III] line, we measure a de-reddened (with E(B-V) = 0.25 mag, \citealp{Gilmozzi2007}) upper limit of $F(H\beta) < 4.5 \times 10^{-17}$ erg/cm$^2$/s, which corresponds to a luminosity of $L(H\beta) < 6.7 \times 10^{29}$ erg/s at the distance of $d = 3.55 \pm 0.77$ kpc. Interestingly, emission due to H$\beta$ from the central WD is clearly detected (see also Figs. \ref{fig:5}, \ref{fig:6}), whose origin is likely due to gas surrounding the progenitor system.  
We can now determine the mass of the western blob using the following formula:

\begin{equation}\label{eq:4}
    M_{ej,b} < 2 \, m_p \, n_e \, V,
\end{equation}

under the assumption that $n_e = n_p = 315.7 \pm 56.4$ cm$^{-3}$, as determined from Eq. \ref{eq:3}, and with $m_p$ being the proton mass. We obtain an upper limit for the bipolar ejecta of $M_{ej,b} < (3.0 \pm 1.0) \times 10^{-6}$ M$_{\odot}$. This estimate does not include any effect of the filling factor due to clumpiness or structured ejecta, suggesting that the upper limit could be also lower up to one order of magnitude. 

We immediately notice that, despite being an upper limit, this value is lower than any measurements reported so far in the literature for the 2011 outburst. Optical studies have reported ejecta masses ranging from $M_{ej}$ = 7 $\times$ 10$^{-6}$ M$_{\odot}$ \citep{Pavana2019}, to $M_{ej}$ = 2 $\times$ 10$^{-5}$ M$_{\odot}$ \citep{Shore2013}, while at other wavelengths we have similar values from X-ray analysis, $M_{ej} \approx$ 10$^{-5}$ M$_{\odot}$ \citep{Chomiuk2014}, and possibly more significant ejecta using radio observations, $M_{ej}$ = 1-30 $\times$ 10$^{-5}$ M$_{\odot}$ \citep{Nelson2014}, while the analysis of the variation of the orbital period before and after the 2011 outburst also suggests a greater ejecta value, $M_{ej}$ = 2 $\times$ 10$^{-5}$ M$_{\odot}$ \citep{Patterson2017}. We note that for the distance of $d = 4.8 \pm 0.5$ kpc \citep{Sokoloski2013}, the upper limit for the mass of the bipolar ejecta is $M_{ej,b,S13} < (6.8 \pm 2.0) \times 10^{-6}$ M$_{\odot}$, a value that is consistent with the estimate given in \citet{Pavana2019}. We also want to remark that this upper limit applies only to the bipolar ejecta identified in the MUSE-NFM observations, and does not take into account the presence of a slower ring component, whose lack of clear evidence in the MUSE-NFM data prevented us from quantifying its contribution to the global ejected mass. A more detailed analysis is still needed, and it will be the topic of a forthcoming publication dedicated to the extended remnant of T Pyx.

\begin{figure}
    \centering
    \includegraphics[width=0.88\linewidth]{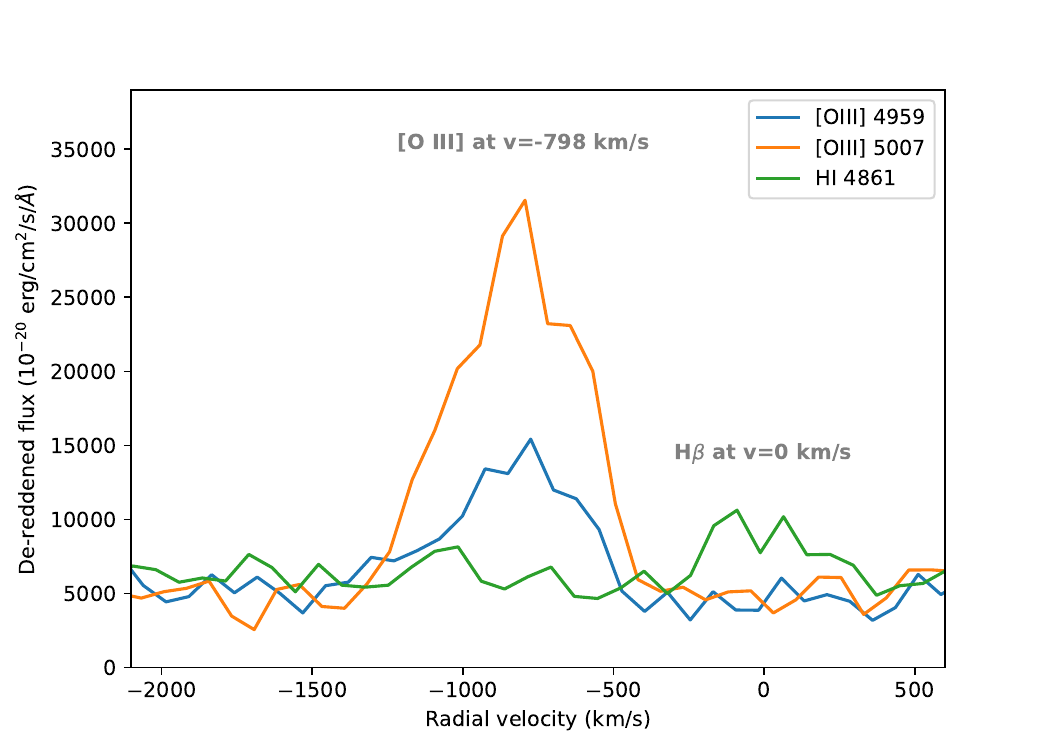}
    \caption{The spectrum of the western blob expressed in radial velocity, with wavelength centered on [O III] 4959 \AA\, (blue data), 5007 \AA\, (orange), and H$\beta$ (green). Note the absence of emission from H$\beta$ at the blue-shift velocity value where emission from [O III] is detected. H$\beta$ emission is however detected at zero radial velocity, whose origin comes likely from the gas surrounding the progenitor WD (work in preparation).}
    \label{fig:6}
\end{figure}

\section{Conclusions}\label{sec:6}

In this work, we have presented a study of the geometrical distribution of the gas surrounding the recurrent nova T Pyx as observed by the ESO/MUSE IFU detector. MUSE has allowed us to perform a spatially-resolved study of the spectral emission from the extended remnant, using the WFM configuration, and the more internal (FoV = 7.5 $\times$ 7.5 arcsec$^2$) regions surrounding the progenitor WD, with an incredible spatial sampling of 0.02 arcsec.  The results of our analysis can be summarized as follows:

\begin{itemize}
    \item The reconstructed Balmer emission line image, using the MUSE-WFM datacube, shows the presence of a tilted ring (see Fig. \ref{fig:2}), showing blue-shifted radial velocities on the Eastern side of the ring, and red-shifted values on the Western side, suggesting the existence of an equatorial ejecta component that has shaped the extended remnant;
    \item We have modeled the reconstructed H$\beta$ remnant emission using a formalism typically used in the study of tilted discs in galaxies and AGN \citep{Begeman1989} but with a radially-expanding velocity for the gas, instead of the generic rotating formalism (see Figs. \ref{fig:3}, \ref{fig:1}). We have measured the inclination of the remnant along our line of sight to be $i = 63.7^{+20.8}_{-13.8}$ deg, while the remnant extension has a radius of $R = 5.6^{+0.7}_{-0.7}$ arcsec, with the ejecta having an expanding velocity of $v_{exp} = 471.7^{+77.2}_{-72.2}$ km/s;
    \item The MUSE-NFM dataset has instead shown the presence of a bipolar outflow characterized by opposite expanding velocities when compared to the extended ring remnant: blue-shifted velocities are observed on the Western ejecta and red-shifted values on the Easter one. This evidence implies the existence of a bipolar outflow, which has been also observed in HST images obtained 3 years after the last outburst, but not before, pointing out that this bipolar outflow originated in the 2011 outburst of T Pyx;
    \item the MUSE-NFM analysis has revealed an offset of the Western blob of $r = 0.69 \pm 0.04 (stat) \pm 0.05 (sys)$ arcsec (see Fig. \ref{fig:4}), which corresponds to an expansion parallax of $\dot{\theta} = (9.4 \pm 0.5) \times 10^{-2}$ arcsec/yr, using the 2014 HST image as reference. With these values we have determined a nebular parallax distance to T Pyx of $d = 3.55 \pm 0.77 $ kpc;
    \item with the knowledge of the distance and the lack of a detected H$\beta$ emission line (see Fig. \ref{fig:6}), we have first estimated the volume of the western blob, and then set an upper limit to the bipolar ejected mass, assuming case B recombination physical conditions for the gas in the ejecta. We have obtained an upper limit of $M_{ej,b} < (3.0 \pm 1.0) \times 10^{-6}$ M$_{\odot}$ for the bipolar ejecta in T Pyx, a value that does not include any correction for the ejecta filling factor. Including this correction, we will get a mass limit that is lower than the total accreted mass on the primary WD during the quiescence phase between the last two recent outbursts of T Pyx \citep{Selvelli2008,Godon2014,Patterson2017}. 
\end{itemize}

In a forthcoming paper, we will discuss in more detail the physical properties of the extended remnant. Our focus will be on examining the impact of multiple outbursts in shaping the observed gas distribution, to facilitate the comparison between the material accreted between outbursts and the average mass of the ejecta. 

\section*{Acknowledgements}

 LI was supported by a VILLUM FONDEN Investigator grant (project number 16599). E.A. acknowledges support by NASA through the NASA Hubble Fellowship grant HST-HF2-51501.001-A awarded by the Space Telescope Science Institute, which is operated by the Association of Universities for Research in Astronomy, Inc., for NASA, under contract NAS5-26555.
 MOH was supported by the Polish National Science Centre grant 2019/32/C/ST9/00577.

\section*{\small Data Availability}
The MUSE datacubes are available upon reasonable request. We plan to upload the datacubes and the codes used for the modeling of the remnant of T Pyx  on a dedicated GitHub repository, which is reported in the manuscript. 




.



\end{document}